\documentclass[prd,aps,nofootinbib,showpacs,preprintnumbers,10pt,twocolumn]{revtex4}
\usepackage{amsmath,amssymb,graphicx,epsfig}

\begin{document}

\preprint{DESY 11-189}

\title{Determining  CP violation angle $\gamma$ with $B$ decays into a scalar/tensor meson}
\author{ Wei Wang \footnote{Email:wei.wang@desy.de}}
\affiliation{Deutsches Elektronen-Synchrotron DESY, D-22607 Hamburg, Germany}

\begin{abstract}
We propose a new way  for determining the CP violation angle $\gamma$ without any hadronic uncertainty. The suggested method is to use the two triangles formed by the decay amplitudes of  $B^\pm\to (D^0,\bar D^0,D_{CP}^0) K^{*\pm}_{0(2)}(1430)$.  The advantages are that  large CP asymmetries are expected in these processes and only singly Cabibbo-suppressed $D$ decay modes are involved.
Measurements of the branching fractions of the neutral $B_d$ decays into $DK^{*}_{0(2)}(1430)$ and  the time-dependent CP asymmetries in    $B_s\to (D^0, \bar D^0) M$ ($M=f_0(980),f_0(1370),  f_2'(1525), f_1(1285), f_1(1420), h_1(1180))$   provide an alternative way to extract the angle $\gamma$, which  will increase  the statistical significance.  
\end{abstract}

\maketitle
 

CP violation in the standard model (SM) originates from a single, irreducible phase in the $3\times 3$ quark mixing matrix called the Cabibbo-Kobayashi-Maskawa (CKM) matrix. Precision test of its unitarity allows us to explore the SM description of the CP violation and reveal  any physics beyond the SM. One of the foremost tasks during the past decades has been to 
study the so-called $(bd)$ unitarity triangle, the graphical
representation of the condition stemming from  the unitarity
of the CKM matrix: $V_{ud}V_{ub}^*+V_{cd}V_{cb}^*+V_{td}V_{tb}^*=0$. 
Its sides can be measured by leptonic and semileptonic meson decays, while  the determinations of the angles  $(\alpha,\beta,\gamma)$, satisfying  $\alpha+\beta+\gamma=180^\circ$,   rely mostly on nonleptonic $B$ decays.

Our knowledge of the angle $\beta$  to a large extent benefits from the gold-plated channel $B\to J/\psi K_S$  and the current results already have a precision  better than $1^\circ$~\cite{Asner:2010qj,CKMfitter}.   The accuracy on the angle
$\alpha$  is around $4^\circ$, thanks to the measurements of  charmless tree dominated processes  $B\to \pi^+\pi^-, B\to  \rho^\pm\pi^\mp, B \to \rho^+\rho^-$ and $B\to a_1^\pm \pi^\mp$. 
In contrast, results for the angle $\gamma$ are less accurate, with a precision of roughly $10^\circ$, which is one of  the main sources of the current uncertainties on the apex of the unitary triangle. 

Since the angle $\gamma\equiv arg(- V_{ud}V_{ub}^*/(V_{cd}V_{cb}^*))$ is the relative weak phase involving the decays $b\to c\bar us$ and $b\to u\bar cs$,  several methods on the basis of the decays $B^\pm\to DK^\pm$, with $D$ being any admixture of $D^0$ and $\bar D^0$, have been proposed (for a review, see Ref.~\cite{Browder:2008em}). The most productive ones are  the Gronau-London-Wyler (GLW) method~\cite{Gronau:1991dp,Gronau:1990ra,Dunietz:1991yd}, with $D$ decaying into the CP eigenstates including $\pi^0 K_S, \pi^+\pi^- K_S, K^+K^-,\pi^+\pi^-$, the Atwood-Dunietz-Soni (ADS) method~\cite{Atwood:1996ci,Atwood:2000ck}, using the Cabibbo-favored and doubly Cabibbo-suppressed $D$ decay modes, and the Giri-Grossman-Soffer-Zupan (GGSZ) method~\cite{Giri:2003ty}, which makes use of a Dalitz-plot distribution of the products of the multi-body $D$ decays.   All  three methods are theoretically clean and do not require any time-dependent measurement. 

In  the GLW method, the sensitivity  of the CP asymmetries to $\gamma$ is proportional to the ratio of the two interfering amplitudes, which is of the order $10\%$. The ADS method demands a detailed knowledge of the doubly Cabibbo-suppressed $D$ decays, while the GGSZ method requires a Dalitz-plot analysis  of multibody $D$ decays. 
In this work, we propose a new method which is based on $B\to D M$ decays with $M$ being a light scalar/tensor meson. The proposed method has both advantages, namely on the one hand the interference and the CP violation in the chosen decay modes are  sizable and on the other hand neither doubly Cabibbo-suppressed $D$ decays nor the Dalitz plot  are needed. 
Among the various $B$ decays into a p-wave scalar/tensor meson to be discussed, of particular interest  are the $B^\pm\to (D^0,\bar D^0,D_{CP}^0) K^{*\pm}_{0(2)}(1430)$ modes,  where $K^*_{0(2)}(1430)$ is a scalar (tensor)  meson with $J^P=0^+(2^+)$. The small (zero) decay constant of $K^*_{0}(1430)$($K^*_{2}(1430)$) compensates the large Wilson coefficient in  the color-allowed amplitude, resulting in similar sizes for the decay amplitudes of $B^\pm\to D^0 K^{*\pm}_{0(2)}(1430)$ and $B^\pm\to\bar D^0 K^{*\pm}_{0(2)}(1430)$. As a consequence, there are large CP asymmetries. Measurements of  branching ratios (BRs) of the neutral $B_d$ decays into $DK^{*}_{0(2)}(1430)$ and  time-dependent CP asymmetries in $B_s\to D M$ ($M=f_0(980),f_0(1370),  $ $ f_2'(1525), f_1(1285), f_1(1420), h_1(1180))$ provide an alternative way to extract the angle $\gamma$. 
For the sake of brevity, hereafter we use $K^*_{0,2}$ and $f_0, f_2'$  to abbreviate $K^*_{0,2}(1430)$ and $f_0(980), f_2'(1525)$, respectively.



\begin{figure}\begin{center}
\includegraphics[scale=0.3]{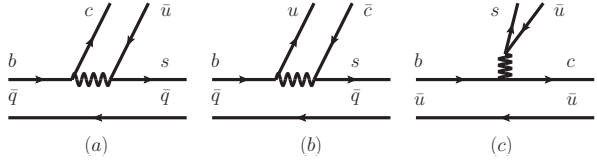}
\caption{Feynman diagrams for the  color-suppressed  contributions in the process $B^-\to D^0 K^{*-}_{0(2)}(1430)$  (a),  $B^-\to \bar D^0 K^{*-}_{0(2)}(1430)$ (b),  and the color-allowed contributions in the  $B^-\to D^0 K^{*-}_{0(2)}(1430)$ (c). In the diagrams (a, b), the spectator quark can also be a $\bar d$ or $\bar s$ quark, in which the light hadron consists of $K^*_0(1430)$, $f_0(980)$ and $f_2'(1525)$.  } \label{fig:feynman}
\end{center}
\end{figure}

All three methods~\cite{Gronau:1991dp,Gronau:1990ra,Dunietz:1991yd,Atwood:1996ci,Atwood:2000ck,Giri:2003ty} to extract $\gamma$ based on $B^\pm \to (D^0, \bar D^0, D_{CP}^0)K^\pm$ use the information that the six decay amplitudes form two triangles in the complex plane,  graphically representing the following identities
\begin{eqnarray}
 \sqrt 2 A(B^+\to D_\pm^0 K^+) = A(B^+ \to D^0K^+) \nonumber\\ \pm A(B^+\to \bar D^0K^+),\nonumber\\
 \sqrt 2 A(B^-\to D_\pm^0 K^-) = A(B^- \to D^0K^-) \nonumber\\ \pm A(B^-\to \bar D^0K^-),\label{eq:identity}
\end{eqnarray} 
where the convention $CP|D^0\rangle =|\bar D^0\rangle$ has been adopted and $D^0_+(D^0_-)$ denotes the CP even (odd) eigenstate.  
The corresponding Feynman diagrams for these processes are given in Fig.~\ref{fig:feynman}.  
Measurements of the decay rates of the six processes completely determine the sides and apexes of the two triangles, 
 in particular the relative phase between $A(B^- \to \bar D^0K^-)$ and $A(B^+ \to D^0K^+)$ is $2\gamma$. 

The shape of the two triangles is governed by two quantities  
\begin{eqnarray}
 r_{B}^{K_J}\equiv\left|{A(B^-\to \bar D^0 {K_J^-})}/{A(B^-\to D^0 K^{-}_J)}\right|,\nonumber\\
 \delta_{B}^{K_J} \equiv arg\left[{e^{i\gamma} A(B^-\to \bar D^0 K^{-}_J)}/{A(B^-\to D^0 K^{-}_J)}\right],\nonumber
\end{eqnarray}
with $K_J= K, K^*_{0,2}$.  The $B^-\to \bar D^0 {K^-}$ is both Cabibbo-suppressed and color suppressed. Thus the ratio  $r_B^K \sim |V_{ub}V_{cs}^*/(V_{cb}V_{us}^*) a_2/a_1|\sim 0.1$ is small and in fact the world averages for the parameters~\cite{CKMfitter}
\begin{eqnarray}
 r_B^K= 0.107\pm 0.010,\;\;
 \delta_B^K=( 112^{+12}_{-13})^\circ \nonumber
\end{eqnarray}
indicate that the two triangles are squashed. Physical observables to be experimentally  measured, defined as
\begin{eqnarray}
 R_{CP\pm}^K &=&2\frac{{\cal B}(B^-\to D_{CP\pm} K^-)+{\cal B}(B^+\to D_{CP\pm} K^+)  }{{\cal B}(B^-\to D^0K^-) +{\cal B}(B^+\to \bar D^0 K^+) }\nonumber\\
&=& 1+(r_{B}^{K})^2\pm 2r_{B}^{K} \cos\delta_{B}^K \cos\gamma,\nonumber\\
 A_{CP\pm}^K &=&\frac{{\cal B}(B^-\to D_{CP\pm} K^-)-{\cal B}(B^+\to D_{CP\pm} K^+)  }{{\cal B}(B^-\to D_{CP\pm} K^-) +{\cal B}(B^+\to D_{CP\pm} K^+) }\nonumber\\
&=&\pm 2r_B^K \sin\delta_{B}^K \sin\gamma /R_{CP\pm}^K, \nonumber
\end{eqnarray}
have a mild sensitivity to the angle $\gamma$, and their values are expected  to  be $ R_{CP\pm}^K \sim 1$ and $ A_{CP\pm}^K\sim 0$.

Since the $K^{*}_{0(2)}$ have the same flavor structure as the $K$ meson, the relations given in Eq.~\eqref{eq:identity} also apply to $B^\pm \to (D^0, \bar D^0, D_{CP}^0) K^{*\pm}_{0(2)}(1430)$.  
We wish to point out that, because of the suppression of the color-allowed decay amplitudes, the low sensitivity problem is highly improved and in particular large CP asymmetries are expected. Although the $K^*_{0(2)}$-emission diagram, as depicted in Fig.~\ref{fig:feynman}(c),  has a large Wilson coefficient $a_1\sim 1$,  the emitted meson is generated by a local vector or  axial-vector current (at the lowest order in $\alpha_s$), whose matrix element between the QCD vacuum and  the $K^*_{0}$($K^*_{2}$) state is small (identically zero).

A crude and model-dependent estimate of  the  amplitudes can be made with the help of the factorization hypothesis
\begin{eqnarray}
&& A(B^-\to \bar D^0 K^{*-}_0) =-V_{ub}V_{cs}^*  C, \nonumber\\
&&A(B^-\to  D^0 K^{*-}_0) = -  V_{cb}V_{us}^* (C-T),\label{eq:factorization}
\end{eqnarray}
where $C= G_F f_{D} a_2(m_{B}^2-m_{K^*_0}^2)  F_0^{B K^*_0}(m_D^2)/\sqrt 2$, $T=G_F f_{K^*_0}a_1(m_{B}^2-m_{D}^2)  F_0^{BD}(m_{K^*_0}^2)/\sqrt 2$, and
$G_F$ is the Fermi constant.  The decay constant, defined via
\begin{eqnarray}
 \langle K^{*-}_0(1430)|\bar s\gamma^\mu u|0\rangle = f_{K^*_0} p_{K^*_0}^\mu,\nonumber
\end{eqnarray}
vanishes in the SU(3) symmetry limit and may get a nonzero but small value due to the symmetry breaking effects. The current experimental data on
$\tau\to K^{*-}_0(1430)\bar\nu_{\tau}$ places an upper bound~\cite{Nakamura:2010zzi}
\begin{eqnarray}
 |f_{K^*_0}|< 107 {\rm MeV}, \nonumber
\end{eqnarray}
which is not very stringent. 
Adopting an estimate based on QCD sum rules~\cite{Cheng:2005nb}
\begin{eqnarray}
 f_{K^*_0}=-24 {\rm MeV}, \;\;\; {\rm or }  \;\;\;
 f_{K^*_0}=36 {\rm MeV}, \nonumber
\end{eqnarray}
which contains a sign ambiguity, we find the relation $2a_1  |f_{K^*_0}|\sim a_2 f_{D}$,
with
the  $D$ meson  decay constant  extracted from $D^-\to \mu\bar\nu_\mu$: $
 f_{D}= (221\pm 18) {\rm MeV}$~\cite{Nakamura:2010zzi}.
Using one set of results for the $B\to K^*_0$ form factors calculated in the perturbative QCD approach~\cite{Li:2008tk} (corresponding to $f_{K^*_0}=36$ MeV), the $B\to D$ form factors from Ref.~\cite{Cheng:2003sm} and $a_2=0.2, a_1=1$
we estimate  $C/T\sim 1.2$  and
\begin{eqnarray}
r_B^{K^*_0}=\left|C{V_{ub} V_{cs}^*}/[{V_{cb}V_{us}^*}(C-T)] \right|\sim 2.
\end{eqnarray}  
The corresponding BRs are  roughly
\begin{eqnarray}
{\cal B}(B^-\to \bar D^0 K^{*-}_0)
\sim 
4\times 10^{-6}.
\end{eqnarray}

\begin{figure}\begin{center}
\includegraphics[scale=0.7]{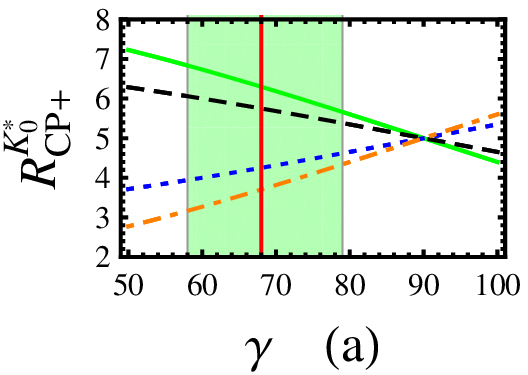}
\includegraphics[scale=0.75]{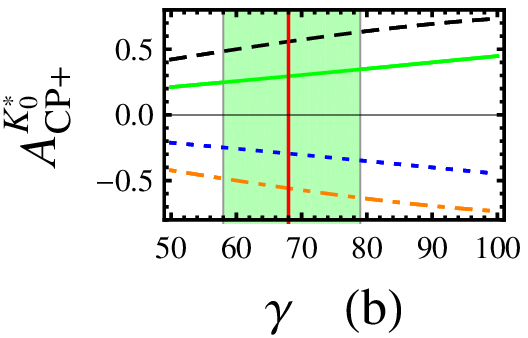}
\includegraphics[scale=0.7]{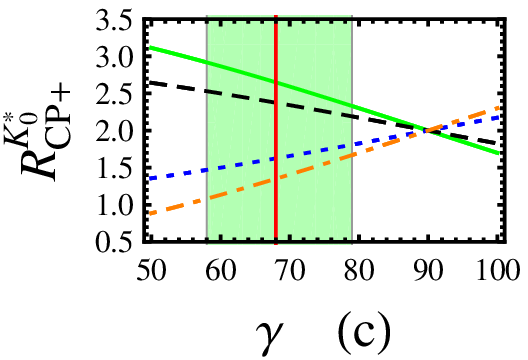}
\includegraphics[scale=0.75]{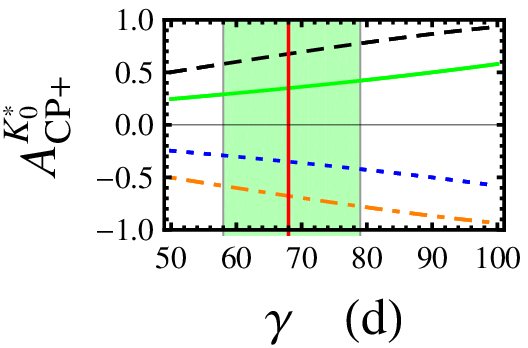}
\includegraphics[scale=0.7]{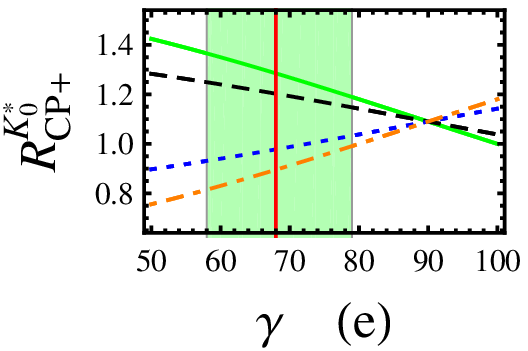}
\includegraphics[scale=0.7]{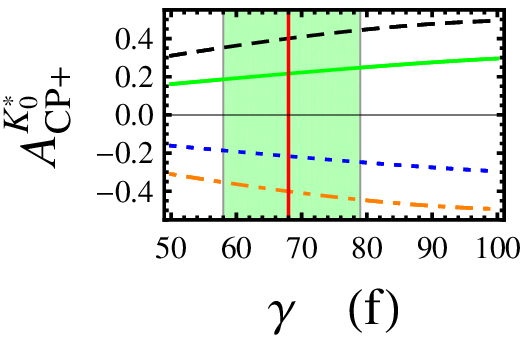}
\caption{ The dependence of   $R_{CP+}^{K^*_0}$ and $A_{CP+}^{K^*_0}$ on $\gamma$. In panels (a,b), $r_{B}^{K^*_0}=2$ is employed,  in panels (c,d) $r_{B}^{K^*_0}=1$ and  in panels (e,f) $r_{B}^{K^*_0}=0.3$. The solid (green), dashed (black), dotted (blue) and dot-dashed (orange) lines in diagrams (a,c,e) correspond to $\delta_{B}^{K^*_0}=(30,60,120,150)^\circ$ respectively,   while the corresponding lines in diagrams (b,d,f) correspond to $\delta_{B}^{K^*_0}=(30,60,-30,-60)^\circ$. The shadowed  (light-green) region denotes the current bounds on $\gamma=(68^{+10}_{-11})^\circ$ from a combined analysis of $B^\pm\to DK^\pm$~\cite{CKMfitter}, in which the vertical (red) line corresponds to the central value.   } \label{fig:dependence}
\end{center}
\end{figure}

Since the strong phase can not be computed at present, we take several benchmark values to illustrate the dependence of $R_{CP+}^{K^*_0}$  and $A_{CP+}^{K^*_0}$ in Fig.~\ref{fig:dependence}. In panels (a,b), $r_{B}^{K^*_0}=2$ is employed, and in panels (c,d) $r_{B}^{K^*_0}=1$. In the last two panels (e,f), we consider the case in which the ratio is not enhanced too much $r_{B}^{K^*_0}=0.3$.
  The solid (green), dashed (black), dotted (blue) and dot-dashed (orange) lines in diagrams (a,c,e) are obtained with $\delta_{B}^{K^*_0}=(30,60,120,150)^\circ$ respectively,   while the corresponding lines in diagrams (b,d,f) correspond to $\delta_{B}^{K^*_0}=(30,60,-30,-60)^\circ$. The shadowed  (light-green) region denotes the current bounds on $\gamma=(68^{+10}_{-11})^\circ$ from a combined analysis of $B^\pm\to DK^\pm$~\cite{CKMfitter}, in which the vertical (red) line corresponds to the central value.  The CP odd quantities can be obtained similarly, for instance $R_{CP-}^{K^*_0}=(R_{CP+}^{K^*_0})_{\delta_{B}^{K^*_0}\to 180^\circ - \delta_{B}^{K^*_0}}$.

Turning to the $B^\pm \to D K_2^{*\pm}$ mode in which the matrix element of the vector and the axial-vector current between the QCD vacuum and the  $K_2^*$ state is zero,  we find a vanishing color-allowed amplitude $T$. Accordingly, the ratio $r_{B}^{K^*_2}$ is from the product of CKM matrix elements which is roughly 0.5. An estimate of the branching ratios can  be made by using the data on the  $B\to J/\psi K_2^*$
\begin{eqnarray}
\frac{{\cal B}( B^-\to D^0 K_2^{*-} )}{{\cal B}(B \to J/\psi K_2^{*0})} &\simeq& x_{K_2^*} \left|
\frac{V_{cb}V_{us}^* }{V_{cb}V_{cs}^*}  \frac{f_{D}}{f_{J/\psi}}\right|^2\sim 0.8\% ,
\end{eqnarray}
with $x_{K_2^*}$ being the ratio  of the form factor products which is evaluated from a recent calculation of $B\to K_2^*$ form factors~\cite{Wang:2010ni}: $x_{K_2^*}\simeq 0.5$. The branching ratio  $
{\cal B}(B \to J/\psi K_2^{*0}) = (4.0\pm 2.4) \times 10^{-4} $~\cite{Burns:2011rn}
 extracted from the data on $B^-\to J/\psi K^-\pi^+\pi^-$~~\cite{:2010if}
gives
\begin{eqnarray}
{\cal B}( B^-\to D^0 K_2^{*-} )\simeq 3\times 10^{-6}.
\end{eqnarray}

The method to use the two triangles formed by the six decay amplitudes for determining $\gamma$ is also valid in  the neutral $B_d$ decays into  $DK_{0,2}^{*0}$, in which the tree amplitude $T$ is identically zero.  The $K^{*}_{0,2}$ is self-tagging, thus no time-dependent measurement is required. Since the amplitudes involving $D^0$ and $\bar D^0$ arise from the same type of diagram, one expects that $\delta_{B}^{K^*_0}\sim 0$. If true,  the CP asymmetries $A_{CP\pm}^{K^*_0}$ would be still close to 0 but $R_{CP\pm}^{K^*_0}$ can largely deviate from 1.

The long-distance contributions in the form of final state interactions (FSI)  might change the factorization analysis in at least two aspects. First, FSI can give   nontrivial strong phases to $C$ and $T$ which are zero in the factorization approach.   Second,  FSI might also modify the size of the amplitudes and the $r_B^{K^*_{0,2}}$. Despite these changes, no hadronic uncertainties will be introduced as the CKM matrix elements in the final state interactions are the same as the ones in Eq.~\eqref{eq:factorization}. To account for such effects, we also show  in Fig.~\ref{fig:dependence}  the dependence of $R_{CP+}^{K^*_0}$ and $A_{CP+}^{K^*_0}$ on $\gamma$ with  different ratios of amplitudes: $r_{B}^{K^*_0}=1$ and $r_{B}^{K^*_0}=0.3$.  The latter corresponds to the sign of  Wilson coefficient $a_2$ reversed namely $a_2=-0.2$.
In this case, despite a small ratio $r_{B}^{K^*_0}=0.3$ the branching fractions ${\cal B}(B^-\to D^0 K^{*-}_{0,2})$ can reach $10^{-5}$.

The $D_{CP}^0$ meson in the final state can be reconstructed in the CP eigenstates, including  the modes $\pi^0 K_S, \pi^+\pi^- K_S, K^+K^-,\pi^+\pi^-$. These modes   have quite large BRs, for instance, ${\cal B}(D^0\to \pi^+\pi^-K_S) \simeq 3\%$~\cite{Nakamura:2010zzi}.   The $K^*_{0,2}$ have significant decay rates into $K\pi$, with ${\cal B}(K^*_0\to K\pi)=(93\pm 10)\%$ and ${\cal B}(K^*_2\to K\pi)=(49.9\pm 1.2)\%$, and the final mesons are also easy to detect in experiments at hadron colliders. 
Moreover, since the CKM matrix elements for the $K^*_0$ and $K_2^*$ are the same,   no knowledge of the resonance structure in this method is required and therefore the angle $\gamma$ can be extracted without any hadronic uncertainty. 
Compared with the BR of $B^-\to \bar D^0 K^-$,  of order $10^{-6}$, which is an unavoidable entry in the currently-adopted methods to determine $\gamma$,  the summed BRs for the channels involving $K^*_0$ and $K_2^*$, of order $10^{-5}$, are comparable or even  larger, and hence their measurements will not be statistically limited.  The large amount of data accumulated by LHCb recently and in future  will lead to a promising prospect of the proposed method.


In the above discussion, we have neglected effects caused by the CP violation in $D$ decays which is anticipated to be small in the standard model. 
Based on the  0.62 fb$^{-1}$ of data  collected  in 2011,  the LHCb collaboration~\cite{LHCb} has  measured 
the difference between  CP asymmetries in singly Cabibbo-suppressed
decays $D^0 \to K^+K^-$ and $D^0 \to \pi^+\pi^-$,
$\Delta A_{CP} \equiv A_{CP}(D^{0}\to K^{+}K^{-})-A_{CP}(D^{0}\to \pi^{+}\pi^{-})$, 
given by
\begin{eqnarray}\label{LHCbData}
\Delta A_{CP} =(-0.82\pm0.21(\text{stat.})\pm0.11(\text{sys.}))\%\,,
\end{eqnarray}
where  the first uncertainty is statistical and the second is systematic.  
Together with    the CDF
results~\cite{CDF2011} and   previous world average from Heavy Flavor Averaging Group~\cite{hfag}, the new world average for $\Delta A_{CP}$  
is found to be~\cite{hfagUpdate}
\begin{eqnarray}\label{NewData}
\Delta A_{CP} = -(0.645\pm 0.180)\%\,.
\end{eqnarray}
Although the new world-averaged $\Delta A_{CP}$ is about $3.6\sigma $ away from zero, its magnitude  is   smaller than 1 percent. 
As a consequence, the CP violation effects in charm decays are  less important  in our method to determine $\gamma$, especially when compared to large uncertainties in current knowledge of $\gamma$. 
Moreover, since the direct CP violation in $D^0\to K^+K^-$ and $D^0\to \pi^+\pi^-$ modes is expected to have opposite signs, part of the CP violation effects will cancel when both decay modes   are used in the reconstruction of $D$ meson.  


Now we turn to the $B_s\to D M$ decays, whose Feynman diagrams are  depicted in Fig.~\ref{fig:feynman} with $\bar q=\bar s$.  It is proposed in Ref.~\cite{Gronau:1990ra,Giri:2001jr}  that the time-dependent CP asymmetries in $B_s\to D\phi$ can  be used to extract  $\gamma$ and this method is applied to a pure annihilation mode $B_s\to D^\pm \pi^\mp$ in Ref.~\cite{Hong:2005vq} and  modes like $B_s\to D \eta(\eta')$ in Ref.~\cite{Chua:2005fj}.  In the example of $B_s\to Df_0$,  there are four decay modes having the amplitudes
\begin{eqnarray}
A(\bar B_s\to \bar  D^0f_0) = V_{ub}V_{cs}^* A_1, 
A(B_s\to D^0f_0) = V_{ub}^*V_{cs}A_1,\nonumber\\
A(\bar B_s\to D^0f_0) = V_{cb}V_{us}^* A_2,
 A( B_s\to \bar D^0f_0) = V_{cb}^*V_{us} A_2.\label{eq:Bsdecayamplitudes}
\end{eqnarray}
For each amplitude, there is only one weak phase in the SM, and therefore no direct CP asymmetry is expected. Any nonzero value from the experiment would be a signal for new physics. 
We define the relative size and strong phase of the two amplitudes as
\begin{eqnarray}
 r_{B_s}^{f_0} = \left|{ V_{ub}V_{cs}^* A_1}/({V_{cb}V_{us}^* A_2})\right|,\;\;\; \delta_{B_s}^{f_0} = arg \left({A_1}/{A_2}\right).\label{eq:Bsratios}
\end{eqnarray}
Since both $A_1$ and $A_2$ are from the same Feynman diagrams, it is likely that $A_1\simeq A_2$, which implies $ r_{B_s}^{f_0} \sim 0.5$ and $\delta_{B_s}^{f_0}\sim 0$. 

The neutral $B_s$ system is described by the mixing 
\begin{eqnarray}
 |B_{L}\rangle = p|B_s^0\rangle + q|\bar B_s^0\rangle, 
 |B_{H}\rangle = p|B_s^0\rangle - q|\bar B_s^0\rangle, \nonumber
\end{eqnarray}
with $|p|^2+|q|^2=1$, and $q/p$ denotes the weak phase in the $B_s-\bar B_s$ mixing 
$
 {q}/{p} = {V_{tb}^* V_{ts}}/({ V_{tb}V_{ts}^*}) 
= e^{-2i\beta_s}.$
In the SM, this ratio is close to unity and the phase $\beta_s$ is negligibly small $\beta_s\simeq -0.019$ rad. The normalized time-dependent decay widths are~\cite{Gronau:1989zb,Aleksan:1990ts}:
\begin{eqnarray}
 &&\Gamma(\bar B_s^0(t) \to D^0(\bar D^0) f_0)= e^{ -t/ \tau_{B_s}}  \Big[1  \nonumber\\
&&+ \cos(\Delta m t)C_{D^0(\bar D^0) f_0} 
+ \sin(\Delta mt)S_{D^0(\bar D^0) f_0}  \Big],
\end{eqnarray}
where $\bar \Gamma$  is the averaged decay width. 
For the corresponding $B_s^0$ decays, the plus signs in front of cosine and sine terms should   be replaced by minus signs.  
Substituting the amplitudes defined in Eq.~\eqref{eq:Bsdecayamplitudes}, we have 
\begin{eqnarray}
 C_{D^0 f_0}  =  C_{\bar D^0 f_0}  = [{1-(r_{B_s}^{f_0})^2}]/[{1+(r_{B_s}^{f_0})^2}],\nonumber\\
 S_{D^0 f_0}  = {-2 r_{B_s}^{f_0}  \sin( \gamma+ \delta_{B_s}^{f_0}+2\beta_s)}/[{1+(r_{B_s}^{f_0})^2}],\nonumber\\ 
 S_{\bar D^0 f_0}  =  {-2 r_{B_s}^{f_0}  \sin( -\gamma+ \delta_{B_s}^{f_0}+2\beta_s)}/[{1+(r_{B_s}^{f_0})^2}].
\end{eqnarray}
The equality $C_{D^0 f_0} = C_{\bar D^0 f_0}$ is a consequence of the uniqueness of the weak phase in decay amplitudes. Since both the strong phase difference $\delta_{B_s}^{f_0}$ and the $B_s-\bar B_s$ mixing phase  are expected small, $S_{D^0 f_0} $ and $S_{\bar D^0 f_0}$ will have similar magnitudes but differ in sign.


The BRs of $\bar B_s\to D f_0(f_2')$ can be estimated by using the experimental data on $B_s\to J/\psi f_0(f_2')$ together with the ratio of the BRs
\begin{eqnarray}
 \frac{{\cal B}(\bar B_s^0\to D^0 f_0)}{{\cal B}(\bar B_s^0\to J/\psi f_0)} &\simeq& x_{B_s}^{f_0}\left|
\frac{V_{cb}V_{us}^* }{V_{cb}V_{cs}^*}  \frac{f_{D}}{f_{J/\psi}}\right|^2\sim (1.3-1.5)\% ,\nonumber\\
 \frac{{\cal B}(\bar B_s^0\to D^0 f_2')}{{\cal B}(\bar B_s^0\to J/\psi f_2')} &\simeq&  x_{B_s}^{f_2'}\left| 
\frac{V_{cb}V_{us}^* }{V_{cb}V_{cs}^*}  \frac{f_{D}}{f_{J/\psi}}\right|^2\sim 0.8\%,\nonumber
\end{eqnarray}
with the product of the form factors $x_{B_s}^{f_0}=(0.8-1.0)$~\cite{Li:2008tk,Colangelo:2010bg}  and $x_{B_s}^{f_2'}=0.50$~\cite{Wang:2010ni}.
A recent measurement~\cite{LHCbBsf2}  
\begin{eqnarray}
{\cal B}(\bar B_s^0\to J/\psi f_0)\sim
{\cal B}(\bar B_s^0\to J/\psi f_2')
\nonumber\\ \sim 0.2 {\cal B}(\bar B_s^0\to J/\psi \phi) 
\sim 2\times 10^{-4} \nonumber
\end{eqnarray}
shows that the $B_s\to D^0 f_0( f_2)$ decays have a BR of order $10^{-6}$. In this estimate the decays $f_0$ into $\pi^+\pi^-$ and $f_2'$ into $K^+K^-$ have been taken into account.

It is straightforward to incorporate the $B_s$ decays into other light p-wave mesons, like $f_0(1370)$, $h_1(1170),  h_1(1380)$, $f_1(1285)$ and $f_1(1420)$. But they  require high statistics to have an impact on $\gamma$, due to either the suppressed production rates in $B_s$ decays~\cite{Li:2009tx} or the difficulty in the reconstruction of the decay modes~\cite{Nakamura:2010zzi}. 

Finally,
we  remark on the BR estimate, which is obtained under the factorization approach in conjunction with the experimental data. The validity of this method can be tested by considering the ratios in the processes $\bar B^0\to (D^0,J/\psi)(\bar K^0,\bar K^{*0})$:
\begin{eqnarray}
 y_{K}\equiv {{\cal B}( \bar B^0\to D^0 \bar K^{0} )}/{{\cal B}(\bar B^0 \to J/\psi \bar K^{0})} \sim 1.4\% ,\nonumber\\
 y_{K^*}\equiv {{\cal B}( \bar B^0\to D^0 \bar K^{*0} )}/{{\cal B}(\bar B^0 \to J/\psi\bar K^{*0})} \sim 0.5\%,
\end{eqnarray}
where the  form factor products are used from Ref.~\cite{Cheng:2003sm}. 
Compared with the data $y_{K}\sim 6.0\%$ and $y_{K^*}\simeq 3.2\%$~\cite{Nakamura:2010zzi},  these ratios are theoretically undershot. If it is the same in  $B/B_s$ decays into $K^*_{0,2}/(f_0,f_2')$,  the estimated BRs will be enhanced roughly by a factor of (4--6),  which makes the proposed method more appealing.



 
In summary, we have explored the possibility to extract the CP violation angle $\gamma$ with $B\to (D^0,\bar D^0,D_{CP}^0) K^{*}_{0(2)}(1430)$ and  $B_s\to (D^0, \bar D^0) M$ ($M=f_0(980),f_0(1370),  f_2'(1525), f_1(1285), f_1(1420), h_1(1180))$. A clean method is to use the two triangles formed by the decay amplitudes of  $B^\pm\to (D^0,\bar D^0,D_{CP}^0) K^{*\pm}_{0(2)}(1430)$.
We expect that $B^\pm\to D^0 K^{*\pm}_{0(2)}(1430)$ and $B^\pm\to\bar D^0 K^{*\pm}_{0(2)}(1430)$ have similar decay rates and the  CP asymmetries have a strong correlation with $\gamma$. Our method does not require the separation of the Cabibbo-suppressed $D$ decays, which are usually buried under the combinatorial background.  With the help of the factorization approach and the relevant experimental data we estimate  the branching ratios of these modes to be of order $10^{-5}- 10^{-6}$.  
Measurements of the branching fractions of $B_d\to DK^{*}_{0(2)}(1430)$ and  time-dependent CP asymmetries in $B_s\to D M$ provide an alternative way to extract the angle $\gamma$.  No knowledge of the resonance structure in this method is required and therefore the angle $\gamma$ can be extracted without any hadronic uncertainty.


{\it Acknowledgement}:
 The author is grateful to A. Ali for carefully reading the manuscript and very useful suggestions, H.N. Li, C.D.  L\"u and Y.M. Wang for useful comments.  
This work is supported by the Alexander von Humboldt Foundation.


\end{document}